\documentclass[%
 reprint,longbibliography,preprintnumbers,
nofootinbib,
 amsmath,amssymb,
 aps
]{revtex4-1}
\pdfoutput=1
\usepackage{graphicx}% Include figure files
\usepackage[utf8]{inputenc}
\usepackage{flushend}
\usepackage{dcolumn}% Align table columns on decimal point
\usepackage{bm}% bold math
\usepackage{balance}
\usepackage[normalem]{ulem}

\usepackage[colorlinks = true,
            linkcolor = blue,
            urlcolor  = blue,
            citecolor =green,
            anchorcolor = blue]{hyperref}
\usepackage{verbatim}
\usepackage{color,ulem}
\usepackage[english]{babel}

\usepackage[utf8]{inputenc}
\input Starburst.fd
\newcommand*\initfamily{\usefont{U}{Starburst}{xl}{n}}\initfamily

\newcommand{\beq}{\begin{eqnarray}}
\newcommand{\eeq}{\end{eqnarray}}
\usepackage{amsmath}
\usepackage{tikz}
\usetikzlibrary{decorations.pathmorphing}
\usetikzlibrary{shapes.misc}
\tikzset{cross/.style={cross out, draw=black, minimum size=8*(#1-\pgflinewidth), inner sep=0pt, outer sep=0pt},
cross/.default={1pt}}
\usetikzlibrary{patterns,math}
\begin{document}

\title{\Large  Quantum confinement theory of the heat capacity of thin films}

\author{\textbf{Alessio Zaccone}$^{1}$}%
 \email{alessio.zaccone@unimi.it}
 
 \vspace{1cm}

\affiliation{$^{1}$Department of Physics ``A. Pontremoli'', University of Milan, via Celoria 16,
20133 Milan, Italy.}

\begin{abstract}
A theory and mechanistic understanding of the thermal properties of solids under nanoscale confinement are currently missing. We develop a theoretical quantum confinement description of thin films which predicts a new physical law for the heat capacity. In particular, due to the suppression of vibrational modes caused by the thin film confinement, the vibrational density of states (VDOS) deviates from the Debye quadratic law in frequency and is, instead, cubic in frequency. This leads to a temperature dependence of the heat capacity which is $\sim T^4$ instead of Debye's $\sim T^3$ law. Furthermore, the new theory predicts a linear increase of the heat capacity upon increasing the nanometric film thickness. Both dependencies are found in excellent agreement with recent experimental data on NbTiN thin films.
\end{abstract}

\maketitle
\section{Introduction}
%\subsection{Debye density of states}
Understanding the thermodynamic properties of nanoscale materials is important for both our fundamental understanding of matter under confinement as well as for a myriad of technological applications. Thin films, in particular, are now routinely being developed with thicknesses in the order of few nanometers, which has a strong impact on their electronic, magnetic and thermal properties. The specific heat or heat capacity is a key thermodynamic property which can be leveraged to infer entropy and enthalpy, and plays an important role for materials used as nanocalorimeters or for thermal rectification and insulation in nanoscale devices \cite{Hellman,Hellman2,Cahill}.
In superconducting materials, such as in Superconducting Nanowire
Single-Photon Detectors (SNSPDs) that find application in
space communications, the phonon heat capacity is a key parameter to control the detection quantum yield \cite{nanowire}.

In spite of intense research, a theoretical mechanistic description of confinement effects leading to the experimentally observed dependence of heat capacity of thin films on temperature and on film thickness is currently missing. Ab initio DFT methods cannot simulate thicknesses beyond few Angstroms \cite{Stengel2006} due to intrinsic limitations in the number of atoms that can be simulated, which is in the order of $10^3$ \cite{Pan2021}, hence insufficient to describe nanometric thin films. 

Recent work has highlighted the existence of a generic or universal effect of thin-film confinement on the propagation of quantum plane waves such as phonons (vibrational excitations) or nearly-free electrons. Without having to assume hard-wall boundary conditions (which are often not realistic in view of the atomic-scale roughness of the film) it has been shown that the occupancy of quasiparticle states in certain regions of momentum space gets suppressed \cite{Phillips,Travaglino_2022,Travaglino_2023}. The formation of unoccupied "hole'' pockets in the k-space distribution of vibrational excitations is responsible for the experimentally observed emergent low-frequency elasticity and viscoelasticity of thin liquid films \cite{PNAS2020,Riedo,Noirez}. This effect is also reflected in the experimentally observed suppression of THz modes in nano-confined water \cite{yang2023suppressed,Yu_2022}. The same effect is responsible for the observed thickness dependence of the dielectric permittivity of ferroelectric thin films \cite{Parker} or the thickness dependence of the Bose-Einstein condensation (BEC) critical temperature in thin films of cold atoms \cite{Travaglino_2022}. In superconducting thin films, the hole pockets in the Fermi sea appear as two symmetric spherical cavities along the confinement axis, which eventually can grow with confinement up to a topological transition point in the Fermi surface that coincides with a point of maximum of the superconducting critical temperature $T_c$ in quantitative agreement with experimental data \cite{Travaglino_2023}.

Here we apply the quantum confinement description of phonons in a thin film and derive a new fundamental law for the heat capacity of thin films that fully takes into account the restriction in the k-space occupancy of phonons due to confinement. The new law is verified in comparison with recent experimental data in terms of both the predicted temperature and thickness dependencies.

\section{Debye theory for bulk solids}
\subsection{Debye density of states}
The number of phonon states with momentum $k'<k$ is given by \cite{Kittel}:
\begin{equation}
N(k'<k) = \frac{V}{(2\pi)^3}Vol_{k}
\end{equation}
which follows from considering that there is one allowed value of momentum $k$ per volume, $(2\pi)^3/V$. By multiplying this by the volume of occupied states in k-space $Vol_{k}$, one then obtains the number of states with $k'<k$ given by the above expression.

For a bulk solid, the volume of occupied states in k-space is given by the Debye sphere: $Vol_{k}=\frac{4}{3}\pi k^{3}$. Hence, the number of states up to momentum $k$ is given by:
\begin{equation}
N(k'<k) = \frac{V}{(2\pi)^3}\frac{4}{3}\pi k^{3}.\label{sphere}
\end{equation}
For a given phonon polarization or branch, the phonon frequency is related to the phonon momentum via: $\omega = v k$. Hence the phonon density of states is given by:
\begin{equation}
g(\omega) = \frac{d}{d\omega} N(\omega'<\omega)=\frac{V}{(2\pi)^3}\frac{4}{3}\pi \frac{d}{d\omega}\left(\frac{\omega}{v}\right)^{3}=\frac{V}{2\pi^2}\frac{\omega^2}{v^3}.
\end{equation}
This is the well-known Debye density of states (VDOS). For a 3D solid, this still has to be multiplied by a factor three to account for the fact that in a solid there are  three polarizations or phonon branches \cite{Kittel}.

\subsection{Debye heat capacity}
The total internal (thermal) energy of the system is given by \cite{Kittel}:
\begin{align}
U&=\int d\omega\, g(\omega)\langle n(\omega)\rangle \hbar \omega\\
&= \int_{0}^{\omega_D}d\omega\left(\frac{3V}{2\pi^2}\frac{\omega^2}{v^3}\right)\left(\frac{\hbar \omega}{e^{\hbar \omega/k_{B}T}-1}\right).\label{energy}
\end{align}
The heat capacity is readily obtained by differentiating the internal energy expression above with respect to temperature:
\begin{equation}
C_{v}=\frac{3V\hbar^{2}}{2\pi^{2}v^{3}k_{B}T^{2}}\int_{0}^{\omega_D}d\omega\frac{\omega^{4}e^{\hbar \omega/k_{B}T}}{\left(e^{\hbar \omega/k_{B}T}-1\right)^{2}}
\end{equation}
Upon changing variable in the ingral, from $\omega$ to $x  = \hbar \omega / k_{B}T$, one thus obtains:
\begin{equation}
C_{v}=9 N k_{B}\left(\frac{T}{\Theta}\right)^{3}\int_{0}^{x_D}dx\frac{x^{4}e^{x}}{\left(e^{x}-1\right)^{2}}
\end{equation}
where $\Theta=\frac{\hbar v}{k_{B}}\left(\frac{6\pi^{2}N}{V}\right)^{1/3}$ is the Debye temperature.

By taking the low-T limit, i.e. the limit $x_D\rightarrow \infty$ in the above integral, one arrives at the following formula \cite{Kittel}:
\begin{equation}
C_v =\frac{12\pi^{4}}{5}N k_{B}\left(\frac{T}{\Theta}\right)^{3}
\end{equation}
with the celebrated $\sim T^{3}$ Debye prediction for the heat capacity of solids.
Upon considering the heat capacity per unit volume, and using the definition of Debye temperature, the following formula is readily retrieved:
\begin{equation}
C_{v,Debye}=\frac{2}{5}\pi^{2}k_{B}^{4}\frac{T^{3}}{\hbar^{3}v^{3}}, \label{Debye}
\end{equation}
where $v$ is the average speed of sound.

\section{Confinement theory for thin films}
\subsection{Density of states under confinement}
\begin{figure}[h]
\centering
\includegraphics[width=\linewidth]{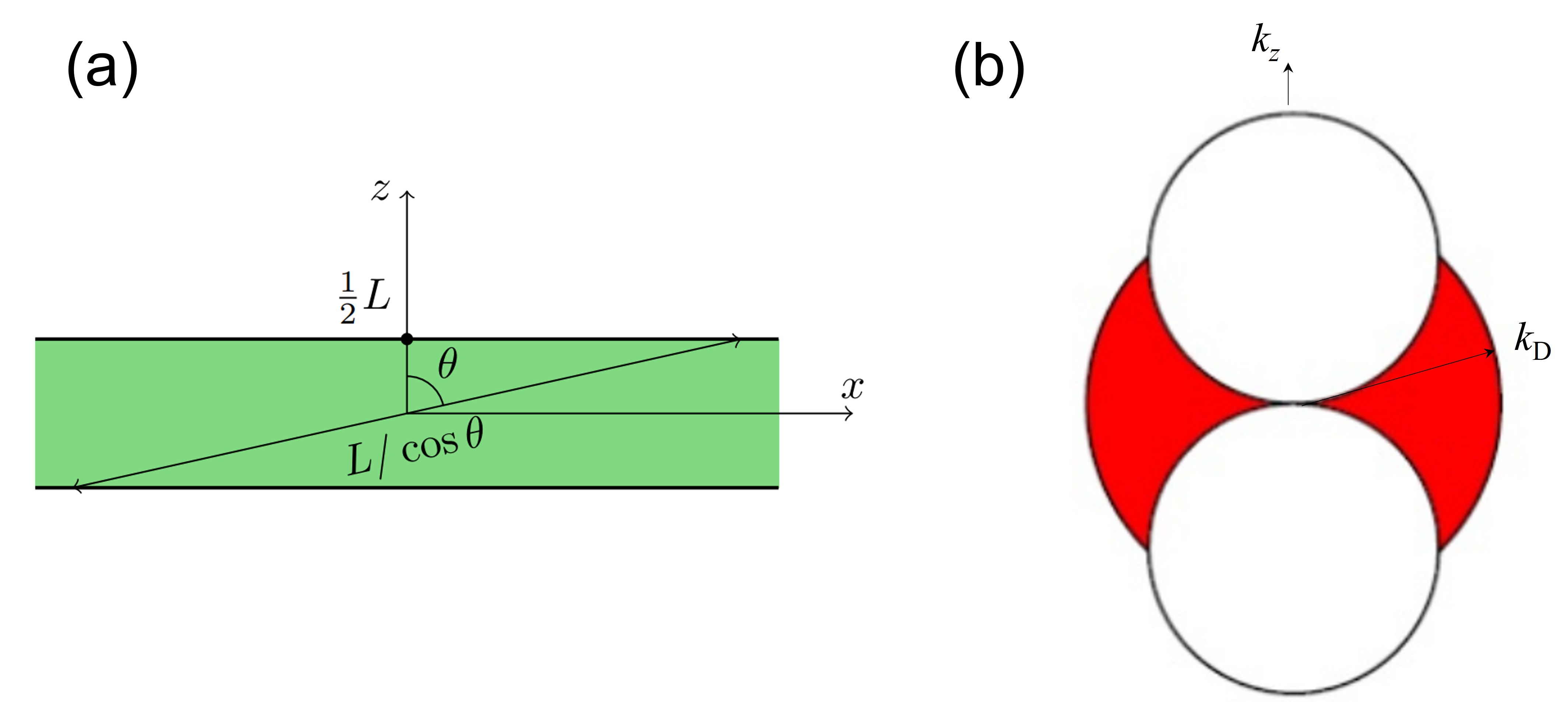}
\caption{Panel (a) shows the thin film geometry in real space (confined along the $z$-axis but unconfined along the $x$ and $y$ axis), with the maximum wavelength that corresponds to a certain polar angle $\theta$. Panel (b) shows a section of the corresponding geometry of $k$-space, where the Debye sphere (of radius $k_{D}$, in red) gets distorted because it now contains two symmetric spheres (hole pockets of radius $\pi/L$, in white) of forbidden states. These are phonon states in $k$-space that remain unoccupied due to confinement along the $z$-axis. See Ref. \cite{Phillips} for a detailed mathematical derivation of this result. }
\label{fig1}
\end{figure}

We now turn to the case of thin film confinement, and we consider a 3D slab which is extended in the horizontal $x$ and $y$ directions, and confined along the vertical $z$ direction, as depicted in Fig. \ref{fig1}(a). The thickness of the film along the confined direction is denoted as $L$.

As demonstrated in previous work \cite{Phillips,Travaglino_2022,Yu_2022}, in the presence of confinement along one direction ($z$-axis), there exists a maximum wavelength, $\lambda_{max}=L/\cos \theta$, where $\theta$ is the polar angle measured with respect to the $z$-axis. This clearly recovers the fact that, in the $xy$ plane orthogonal to $z$, there is no confinement for in-plane phonon propagation, while, along $z$, the maximum wavelength is set by the film thickness $L$. This means that, for each orientation $\theta$, there exists a minimum value of the phonon momentum set by confinement:
\begin{equation}
k_{min}=2 \pi \cos \theta / L.\label{parametric}
\end{equation}
This is the parametric equation of two identical spheres which are the mirror-image of each other across the $xy$ plane, as predicted in Ref. \cite{Phillips} and shown here in Fig. \ref{fig1}(b).
The two ``hollow'' spheres, which are contained within the Debye sphere, represent the forbidden states (hole pockets) that cannot be occupied due to the confinement. Upon further increasing the confinement (i.e. decreasing $L$) these two forbidden spheres eventually touch the Debye sphere surface and, for further confinement, the surface of the highest momentum is no longer spherical, as depicted schematically in Fig. \ref{fig1}(b). 

In the latter case, the occupied volume in k-space is no longer given by a sphere as in Eq. \eqref{sphere}, but it is given instead by the volume of the Debye sphere minus its intersection with the volumes of the two spheres of forbidden states defined by Eq. \eqref{parametric}. The intersection volume of the $L$-dependent volume of the two white spheres in Fig. \ref{fig1} with the Debye red sphere, denoted as $V_{inter}$, can be exactly calculated using Eq. \eqref{parametric} and standard tools of solid geometry. This is done by integrating the areas of stacked disks along the $z$-direction (in general, the volume of a
cylindrically symmetric object is obtained by summing the areas of all the
disks that are stacked on each other to form the object;
since these disks are densely infinite, the sum is, in fact, an integral) \cite{Travaglino_2023}. The exact calculation gives $V_{inter}=\frac{4\pi k^3}{3}-\frac{L k^4}{2}$ and, thus, $Vol_{k}=\frac{4\pi k^3}{3}-V_{inter}$, follows as \cite{Travaglino_2022,Travaglino_2023}:
\begin{equation}
Vol_{k}=\frac{L k^{4}}{2}.
\end{equation}
The number of states in k-space with $k<k'$ then readily follows as:
\begin{equation}
N(k<k') = \frac{V}{(2\pi)^{3}}\frac{L k^{4}}{2}.
\end{equation}
The phonon density of states in the thin film then follows immediately:
\begin{equation}
g(\omega) = \frac{d}{d\omega} N(\omega<\omega')=\frac{V}{4\pi^{3}} L  \frac{\omega^{3}}{v^4}\label{VDOS}
\end{equation}
which exhibits a cubic frequency dependence $\sim \omega^{3}$ that was verified both experimentally (inelastic neutron scattering) and by molecular dynamics simulations in Ref. \cite{Yu_2022}. Importantly, the $\omega^3$ law holds for both crystalline thin films as well as for completely amorphous thin films.

As before, the above VDOS is for just one phonon polarization, and a factor of three has to be implemented when computing the total internal energy $U$ \cite{Kittel}.

Also to be noted, is the dependence of the VDOS on the film thickness $L$, and the inverse quartic dependence on the average speed of sound $v$.

As already pointed out in Ref. \cite{Yu_2022}, the above VDOS crosses over into the Debye VDOS when the frequency becomes $\omega_{\times}=\frac{2\pi}{L}v$. 
Hence, the full VDOS can be written as follows:
\begin{equation}
\begin{split}
g(\omega)&=\frac{V}{4\pi^{3}} L  \frac{\omega^{3}}{v^4}~~~\text{for}~~~ \omega < \omega_{\times},\\
g(\omega)&=\frac{V}{2\pi^2}\frac{\omega^2}{v^3}~~~\text{for}~~~ \omega > \omega_{\times}.
\end{split} \label{full}
\end{equation}

This crossover was verified experimentally and numerically in Ref. \cite{Yu_2022}.
For a thin film which is several nanometers thick, the crossover frequency $\omega_{\times}$ is on the order of few TeraHertz (THz), hence about less than one order of magnitude smaller than the Debye frequency  (which is the highest frequency of the system). Hence $\omega_{\times}$ is a quite high frequency value such that it is safe to take the heat capacity integral up to $\omega_{\times}$ instead of up to $\omega_{D}$. This tantamounts to ignoring the additional Debye contribution that is brought about by the integral from $\omega_{\times}$ to $\omega_{D}$, since this latter integral is expected to be much smaller than the integral from $0$ to $\omega_{\times}$. Also, as shown below, this contribution vanishes anyway upon taking the low-$T$ limit.\\

\subsection{Heat capacity of thin films}
To derive the heat capacity, one proceeds \emph{mutatis mutandis} in the same way as for the Debye theory, i.e. by using the new VDOS Eq. \eqref{VDOS} inside the internal energy formula Eq. \eqref{energy},
\begin{align}
U&=\int d\omega\, g(\omega)\langle n(\omega)\rangle \hbar \omega\\
&= \int_{0}^{\omega_{\times}}d\omega\left(\frac{3V}{4\pi^{3}} L  \frac{\omega^{3}}{v^4}\right)\left(\frac{\hbar \omega}{e^{\hbar \omega/k_{B}T}-1}\right).\label{energy_thin}
\end{align}

Upon differentiating with respect to $T$, the heat capacity per unit volume is obtained as:
\begin{align}
C_{v}&=\frac{3}{4\pi^{3}}\frac{L}{v^{4}}\frac{\hbar^2}{k_{B}T^{2}}\int_{0}^{\omega_{\times}}d\omega\frac{\omega^{5}e^{\hbar \omega/k_{B}T}}{\left(e^{\hbar \omega/k_{B}T}-1\right)^{2}}\\
&=\frac{3 k_B L}{4\pi^{3}}\left(\frac{k_B T}{\hbar v}\right)^{4}\int_{0}^{\infty}dx\frac{x^{5}e^{ x}}{\left(e^{x}-1\right)^{2}}. \label{conf}
\end{align}
In the integral we took the upper limit as $\omega_{\times}$ instead of $\omega_{D}$ because, as explained above, $\omega_{\times}$ is just a fraction of $\omega_{D}$ and anyway this difference disappears when taking the low-temperature limit, $x_{\times}=\hbar\omega_{\times}/k_{B}T \rightarrow \infty$, in the second equality. Furthermore, the integral is correctly extended only up to $\omega_{\times}$ because that is the frequency which separates the regime where the VDOS is given by $\omega^3$ from the regime where the VDOS is given by $\omega^2$, as found quantitatively on the basis of simulation as and experiments in \cite{Yu_2022}.

The integral in $x$ can be evaluated as:
\begin{equation}
\int_{0}^{\infty}dx\frac{x^{5}e^{ x}}{\left(e^{x}-1\right)^{2}}=120\, \zeta(5) \approx 124.431
\end{equation}
where $\zeta$ denotes the Riemann zeta function. 

We thus finally arrive at the formula for the heat capacity of thin films:
\begin{equation}
C_{v}=120\, \zeta(5)\frac{3}{4\pi^{3}}\frac{L}{v^{4}}k_{B}\left(\frac{k_{B}T}{\hbar}\right)^{4}.\label{key_result}
\end{equation}
This is the most important result of this paper, and represents an altogether new fundamental physical law.
This formula has the correct physical units, as one can verify. It exhibits a new $\sim T^{4}$ dependence of the heat capacity on temperature, and a new dependence $\sim L$ on the film thickness. This is a new law of physics brought about by 
quantum confinement of phonons, which we can now compare to recent experimental observations.

\subsection{Recovering the Debye formula in the $L\rightarrow \infty$ limit}
We also observe that in the large $L$ limit the above expression correctly reduces to the Debye expression for the bulk material.
This fact can be shown as follows. 
Using the full VDOS Eq. \eqref{full} in the first of Eq. \eqref{energy_thin}, one obtains:
\begin{equation}
\begin{split}
C_v(T,L)=& C_{v,Debye} + \frac{3 k_B L}{4\pi^{3}}\left(\frac{k_B T}{\hbar v}\right)^{4}\int_{0}^{x_{\times}}dx\frac{x^{5}e^{ x}}{\left(e^{x}-1\right)^{2}}  \\
& - \frac{3 k_B}{2\pi^2 } \left(\frac{k_B T}{\hbar v}\right)^3 \int_{0}^{x_{\times}}dx \frac{x^{4}e^{ x}}{\left(e^{x}-1\right)^{2}}, 
\end{split}
\end{equation}
where $C_{v,Debye}$ denotes the Debye expression Eq. \eqref{Debye} and $x_{\times} =\frac{2\pi v}{L k_B T}$. At low temperature and for thin films, one immediately recovers Eq. \eqref{conf}.
Upon rearranging, we obtain:
\begin{equation}
\frac{C_v(T,L)}{C_{v,Debye}}=1 + \frac{15 L}{8 \pi^5} \frac{k_B T}{\hbar v} \int_{0}^{x_{\times}}\frac{x^5 e^x}{(e^x -1)^2}dx.
\end{equation}
Therefore, in the limit $L \rightarrow \infty$, we have that $x_{\times} \rightarrow 0$ and the integral in the above equation shrinks to zero leaving the Debye formula. This shows the ability of the new formula for thin films to correctly recover the bulk results in the appropriate $L \rightarrow \infty$ limit.

\section{Comparison with experimental data}
\begin{figure}[h]
\centering
\includegraphics[width=\linewidth]{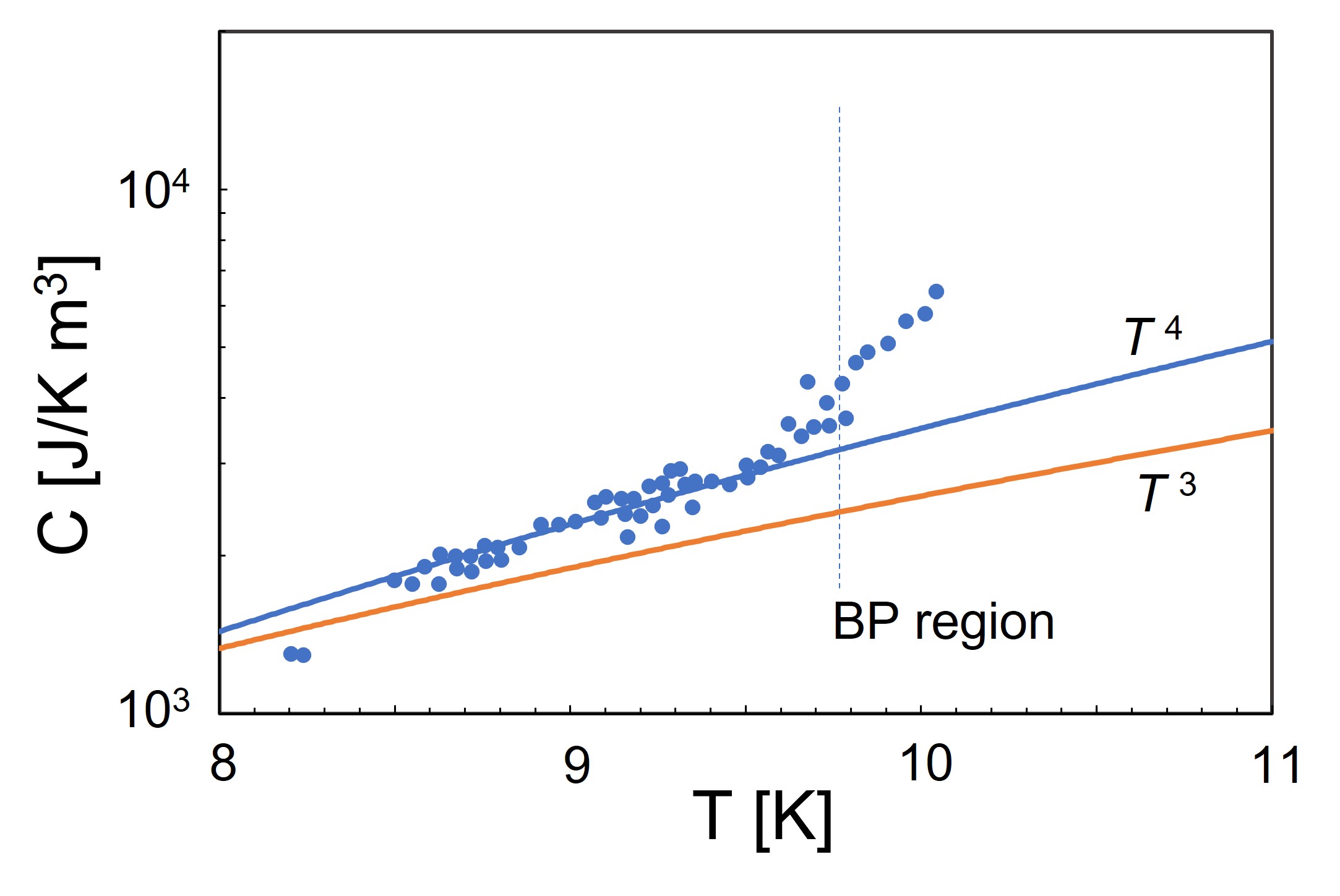}
\caption{Comparison between the $T^4$ temperature dependence of the phonon heat capacity predicted by Eq. \eqref{key_result} (blue line) and experimental data (circles) on NbTiN thin films ($L =6$ nm) from Ref. \cite{Sidorova_2023}. There is only one parameter in the comparison, which is the speed of sound $v \sim 4000$ m/s, which has been taken as a characteristic value of speed of sound in metals and a plausible value for this material.
The Debye $T^3$ scaling (orange line) is also shown for reference. The dashed line indicates the onset of deviation from the $T^4$ scaling due to the boson peak anomaly which in amorphous thin films typically shows up for $T >10$ K \cite{Parker}.}
\label{fig2}
\end{figure}

In Fig. \ref{fig2} we present a comparison with recent experimental data for the phonon heat capacity of NbTiN amorphous films as a function of temperature, taken from Ref. \cite{Sidorova_2023}. As shown in the comparison, the experimental data follow the new $T^4$ law for the heat capacity, instead of the Debye $T^3$ valid for the bulk material.

The experimental data are in good agreement with the $T^4$ law predicted by Eq. \eqref{key_result} up to $T \approx 10$ K where the heat capacity of amorphous thin films is known to be affected by the boson peak (BP) anomaly, which shows up as a peak in the Debye-normalized heat capacity \cite{Parker,Wei-Hua}, and originates from anharmonic damping of acoustic phonons \cite{Baggioli}.

\begin{figure}[h]
\centering
\includegraphics[width=0.9\linewidth]{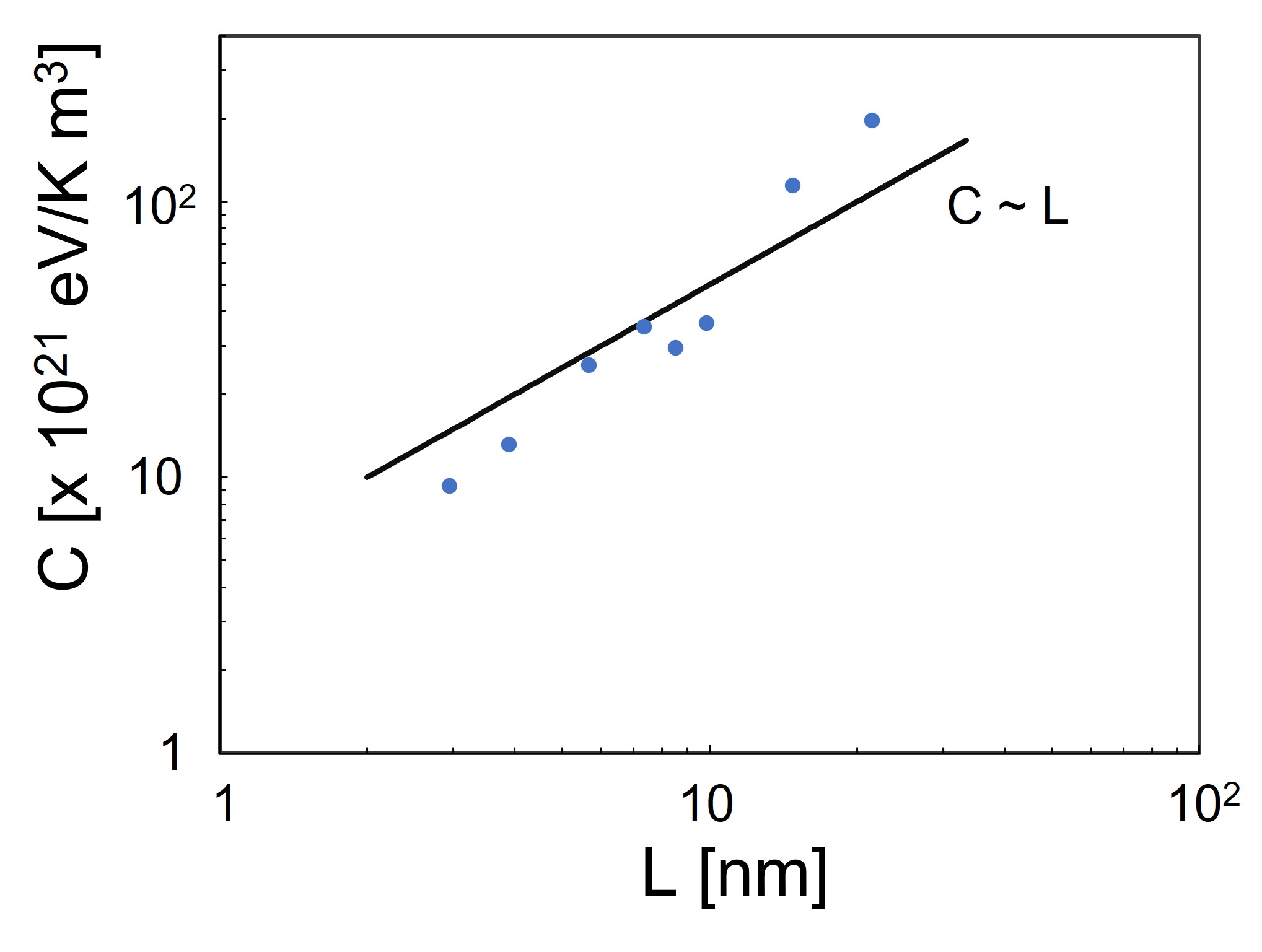}
\caption{Comparison between the $C \propto L$ thickness dependence of the heat capacity predicted by Eq. \eqref{key_result} (solid line) and experimental data (circles) on NbN thin films ($L =6$ nm) from Ref. \cite{Sidorova_2020}. The experimental data were taken at temperatures about $10$K or slightly below.}
\label{fig3}
\end{figure}

As opposed to the Debye theory, which obviously does not feature any dependence on the system length-scale, the new law Eq. \eqref{key_result} shows a linear increasing dependence of the heat capacity on the film thickness, $C \sim L$.
We now test also this prediction against experimental measurements for the same system, i.e. thin films of NbN, from Ref. \cite{Sidorova_2020}.
The comparison is shown in Fig. \ref{fig3}. While the data could also support a dependence on $L$ with a power exponent slightly larger than one, the experimental noise is such that a linear dependence may well be within the error bar.

We note that this linearly increasing trend of the heat capacity with the film thickness $L$ has been reported experimentally also in previous work, notably for aluminum thin films in the range $13.5$ nm to $200$ nm \cite{Song_linear}. This fact reflects the universal character of this fundamental new law, which is independent of the chemical composition of the material, and is induced by the generic confinement physics of vibrational modes. 

This theory also offers a mechanistic understanding of this system-size dependence of the heat capacity: upon increasing the confinement, the hole pockets inside the Debye sphere become larger, which highlights the suppression of vibrational modes in the acoustic to THz vibrational spectrum \cite{Yu_2022,yang2023suppressed} and the redistribution of modes in k-space on a larger and distorted Debye surface (Fig. \ref{fig1}(b)). As a consequence, at low temperature the density of vibrational modes available to efficiently store heat as internal energy is lower, which thus decreases the heat capacity of the material upon decreasing the film thickness.

\section{Conclusions}
In summary, we have derived an analytical formula for the heat capacity of thin films based on a quantum confinement model for vibrational modes. Due to the nanometric confinement along one spatial direction, there exist spherical hole pockets inside the Debye sphere (Fig. \ref{fig1}(b)), which are associated with suppressed modes in the acoustic to THz regime of the vibrational spectrum. This, in turn, gives rise to a redistribution of modes on the distorted Debye surface, with a modified $\sim \omega^3$ density of states that increases more steeply compared to the Debye $\omega^2$ density of states. This, in turn, gives rise to a new $T^4$ law for the heat capacity (in lieu of the $T^3$ Debye law), which we have verified here in comparison with recent experimental data on NbTiN thin films from the literature. The new law also predicts that the heat capacity increases linearly with the film thickness $L$, which is again in agreement with the experimentally observed behaviour. 

In future work, this theory can be extended in several directions such as the inclusion of phonon transmission factors between the thin film and the substrates, and the computation of quantum yields for single-photon detection of superconducting thin films. The theory can be extended to account for the effect of structural disorder, by including the boson peak effect \cite{zaccone2023}.
Also, it will be interesting to apply the theory to more exotic materials such as few-layer quasi-2D van der Waals materials, e.g. 
graphene, BN, MoS$_2$, and amorphous and crystalline membranes. In this perspective, it will be important to start from a theory of layered materials for the VDOS and then apply the confinement concept to arrive at a description of few-layer materials.

\subsection*{Acknowledgments} 
A.Z. gratefully acknowledges funding from the European Union through Horizon Europe ERC Grant number: 101043968 ``Multimech'', and from US Army Research Office through contract nr. W911NF-22-2-0256. Several discussions with Dr. Mariia Sidorova and Dr. Alexei Semenov (DLR Berlin) are gratefully acknowledged. In particular, I am indebted to Dr. Alexei Semenov for pointing out the manipulation leading to Eq. (22).
\bibliographystyle{apsrev4-1}

\bibliography{refs}

\begin{thebibliography}{22}%
\makeatletter
\providecommand \@ifxundefined [1]{%
 \@ifx{#1\undefined}
}%
\providecommand \@ifnum [1]{%
 \ifnum #1\expandafter \@firstoftwo
 \else \expandafter \@secondoftwo
 \fi
}%
\providecommand \@ifx [1]{%
 \ifx #1\expandafter \@firstoftwo
 \else \expandafter \@secondoftwo
 \fi
}%
\providecommand \natexlab [1]{#1}%
\providecommand \enquote  [1]{``#1''}%
\providecommand \bibnamefont  [1]{#1}%
\providecommand \bibfnamefont [1]{#1}%
\providecommand \citenamefont [1]{#1}%
\providecommand \href@noop [0]{\@secondoftwo}%
\providecommand \href [0]{\begingroup \@sanitize@url \@href}%
\providecommand \@href[1]{\@@startlink{#1}\@@href}%
\providecommand \@@href[1]{\endgroup#1\@@endlink}%
\providecommand \@sanitize@url [0]{\catcode `\\12\catcode `\$12\catcode `\&12\catcode `\#12\catcode `\^12\catcode `\_12\catcode `\%12\relax}%
\providecommand \@@startlink[1]{}%
\providecommand \@@endlink[0]{}%
\providecommand \url  [0]{\begingroup\@sanitize@url \@url }%
\providecommand \@url [1]{\endgroup\@href {#1}{\urlprefix }}%
\providecommand \urlprefix  [0]{URL }%
\providecommand \Eprint [0]{\href }%
\providecommand \doibase [0]{http://dx.doi.org/}%
\providecommand \selectlanguage [0]{\@gobble}%
\providecommand \bibinfo  [0]{\@secondoftwo}%
\providecommand \bibfield  [0]{\@secondoftwo}%
\providecommand \translation [1]{[#1]}%
\providecommand \BibitemOpen [0]{}%
\providecommand \bibitemStop [0]{}%
\providecommand \bibitemNoStop [0]{.\EOS\space}%
\providecommand \EOS [0]{\spacefactor3000\relax}%
\providecommand \BibitemShut  [1]{\csname bibitem#1\endcsname}%
\let\auto@bib@innerbib\@empty
%</preamble>
\bibitem [{\citenamefont {Queen}\ and\ \citenamefont {Hellman}(2009)}]{Hellman}%
  \BibitemOpen
  \bibfield  {author} {\bibinfo {author} {\bibfnamefont {D.~R.}\ \bibnamefont {Queen}}\ and\ \bibinfo {author} {\bibfnamefont {F.}~\bibnamefont {Hellman}},\ }\href {\doibase 10.1063/1.3142463} {\bibfield  {journal} {\bibinfo  {journal} {Review of Scientific Instruments}\ }\textbf {\bibinfo {volume} {80}},\ \bibinfo {pages} {063901} (\bibinfo {year} {2009})},\ \Eprint {http://arxiv.org/abs/https://pubs.aip.org/aip/rsi/article-pdf/doi/10.1063/1.3142463/13524164/063901\_1\_online.pdf} {https://pubs.aip.org/aip/rsi/article-pdf/doi/10.1063/1.3142463/13524164/063901\_1\_online.pdf} \BibitemShut {NoStop}%
\bibitem [{\citenamefont {Revaz}\ \emph {et~al.}(2005)\citenamefont {Revaz}, \citenamefont {Zink},\ and\ \citenamefont {Hellman}}]{Hellman2}%
  \BibitemOpen
  \bibfield  {author} {\bibinfo {author} {\bibfnamefont {B.}~\bibnamefont {Revaz}}, \bibinfo {author} {\bibfnamefont {B.}~\bibnamefont {Zink}}, \ and\ \bibinfo {author} {\bibfnamefont {F.}~\bibnamefont {Hellman}},\ }\href {\doibase https://doi.org/10.1016/j.tca.2005.04.004} {\bibfield  {journal} {\bibinfo  {journal} {Thermochimica Acta}\ }\textbf {\bibinfo {volume} {432}},\ \bibinfo {pages} {158} (\bibinfo {year} {2005})},\ \bibinfo {note} {thermodynamics and calorimetry of thin films: Papers presented at the 8th Lähnwitzseminar on Calorimetry}\BibitemShut {NoStop}%
\bibitem [{\citenamefont {Oh}\ \emph {et~al.}(2010)\citenamefont {Oh}, \citenamefont {Ko}, \citenamefont {Ramanathan},\ and\ \citenamefont {Cahill}}]{Cahill}%
  \BibitemOpen
  \bibfield  {author} {\bibinfo {author} {\bibfnamefont {D.-W.}\ \bibnamefont {Oh}}, \bibinfo {author} {\bibfnamefont {C.}~\bibnamefont {Ko}}, \bibinfo {author} {\bibfnamefont {S.}~\bibnamefont {Ramanathan}}, \ and\ \bibinfo {author} {\bibfnamefont {D.~G.}\ \bibnamefont {Cahill}},\ }\href {\doibase 10.1063/1.3394016} {\bibfield  {journal} {\bibinfo  {journal} {Applied Physics Letters}\ }\textbf {\bibinfo {volume} {96}},\ \bibinfo {pages} {151906} (\bibinfo {year} {2010})},\ \Eprint {http://arxiv.org/abs/https://pubs.aip.org/aip/apl/article-pdf/doi/10.1063/1.3394016/13168592/151906\_1\_online.pdf} {https://pubs.aip.org/aip/apl/article-pdf/doi/10.1063/1.3394016/13168592/151906\_1\_online.pdf} \BibitemShut {NoStop}%
\bibitem [{\citenamefont {You}\ \emph {et~al.}(2018)\citenamefont {You}, \citenamefont {Quan}, \citenamefont {Wang}, \citenamefont {Ma}, \citenamefont {Yang}, \citenamefont {Liu}, \citenamefont {Li}, \citenamefont {Li}, \citenamefont {Wang}, \citenamefont {Liang}, \citenamefont {Wang},\ and\ \citenamefont {Xie}}]{nanowire}%
  \BibitemOpen
  \bibfield  {author} {\bibinfo {author} {\bibfnamefont {L.}~\bibnamefont {You}}, \bibinfo {author} {\bibfnamefont {J.}~\bibnamefont {Quan}}, \bibinfo {author} {\bibfnamefont {Y.}~\bibnamefont {Wang}}, \bibinfo {author} {\bibfnamefont {Y.}~\bibnamefont {Ma}}, \bibinfo {author} {\bibfnamefont {X.}~\bibnamefont {Yang}}, \bibinfo {author} {\bibfnamefont {Y.}~\bibnamefont {Liu}}, \bibinfo {author} {\bibfnamefont {H.}~\bibnamefont {Li}}, \bibinfo {author} {\bibfnamefont {J.}~\bibnamefont {Li}}, \bibinfo {author} {\bibfnamefont {J.}~\bibnamefont {Wang}}, \bibinfo {author} {\bibfnamefont {J.}~\bibnamefont {Liang}}, \bibinfo {author} {\bibfnamefont {Z.}~\bibnamefont {Wang}}, \ and\ \bibinfo {author} {\bibfnamefont {X.}~\bibnamefont {Xie}},\ }\href {\doibase 10.1364/OE.26.002965} {\bibfield  {journal} {\bibinfo  {journal} {Opt. Express}\ }\textbf {\bibinfo {volume} {26}},\ \bibinfo {pages} {2965} (\bibinfo {year} {2018})}\BibitemShut {NoStop}%
\bibitem [{\citenamefont {Stengel}\ and\ \citenamefont {Spaldin}(2006)}]{Stengel2006}%
  \BibitemOpen
  \bibfield  {author} {\bibinfo {author} {\bibfnamefont {M.}~\bibnamefont {Stengel}}\ and\ \bibinfo {author} {\bibfnamefont {N.~A.}\ \bibnamefont {Spaldin}},\ }\href {\doibase 10.1038/nature05148} {\bibfield  {journal} {\bibinfo  {journal} {Nature}\ }\textbf {\bibinfo {volume} {443}},\ \bibinfo {pages} {679} (\bibinfo {year} {2006})}\BibitemShut {NoStop}%
\bibitem [{\citenamefont {Pan}(2021)}]{Pan2021}%
  \BibitemOpen
  \bibfield  {author} {\bibinfo {author} {\bibfnamefont {J.}~\bibnamefont {Pan}},\ }\href {\doibase 10.1038/s43588-021-00034-x} {\bibfield  {journal} {\bibinfo  {journal} {Nature Computational Science}\ }\textbf {\bibinfo {volume} {1}},\ \bibinfo {pages} {95} (\bibinfo {year} {2021})}\BibitemShut {NoStop}%
\bibitem [{\citenamefont {Phillips}\ \emph {et~al.}(2021)\citenamefont {Phillips}, \citenamefont {Baggioli}, \citenamefont {Sirk}, \citenamefont {Trachenko},\ and\ \citenamefont {Zaccone}}]{Phillips}%
  \BibitemOpen
  \bibfield  {author} {\bibinfo {author} {\bibfnamefont {A.~E.}\ \bibnamefont {Phillips}}, \bibinfo {author} {\bibfnamefont {M.}~\bibnamefont {Baggioli}}, \bibinfo {author} {\bibfnamefont {T.~W.}\ \bibnamefont {Sirk}}, \bibinfo {author} {\bibfnamefont {K.}~\bibnamefont {Trachenko}}, \ and\ \bibinfo {author} {\bibfnamefont {A.}~\bibnamefont {Zaccone}},\ }\href {\doibase 10.1103/PhysRevMaterials.5.035602} {\bibfield  {journal} {\bibinfo  {journal} {Phys. Rev. Mater.}\ }\textbf {\bibinfo {volume} {5}},\ \bibinfo {pages} {035602} (\bibinfo {year} {2021})}\BibitemShut {NoStop}%
\bibitem [{\citenamefont {Travaglino}\ and\ \citenamefont {Zaccone}(2022)}]{Travaglino_2022}%
  \BibitemOpen
  \bibfield  {author} {\bibinfo {author} {\bibfnamefont {R.}~\bibnamefont {Travaglino}}\ and\ \bibinfo {author} {\bibfnamefont {A.}~\bibnamefont {Zaccone}},\ }\href {\doibase 10.1088/1361-6455/ac5583} {\bibfield  {journal} {\bibinfo  {journal} {Journal of Physics B: Atomic, Molecular and Optical Physics}\ }\textbf {\bibinfo {volume} {55}},\ \bibinfo {pages} {055301} (\bibinfo {year} {2022})}\BibitemShut {NoStop}%
\bibitem [{\citenamefont {Travaglino}\ and\ \citenamefont {Zaccone}(2023)}]{Travaglino_2023}%
  \BibitemOpen
  \bibfield  {author} {\bibinfo {author} {\bibfnamefont {R.}~\bibnamefont {Travaglino}}\ and\ \bibinfo {author} {\bibfnamefont {A.}~\bibnamefont {Zaccone}},\ }\href {\doibase 10.1063/5.0132820} {\bibfield  {journal} {\bibinfo  {journal} {Journal of Applied Physics}\ }\textbf {\bibinfo {volume} {133}},\ \bibinfo {pages} {033901} (\bibinfo {year} {2023})},\ \Eprint {http://arxiv.org/abs/https://pubs.aip.org/aip/jap/article-pdf/doi/10.1063/5.0132820/16766705/033901\_1\_online.pdf} {https://pubs.aip.org/aip/jap/article-pdf/doi/10.1063/5.0132820/16766705/033901\_1\_online.pdf} \BibitemShut {NoStop}%
\bibitem [{\citenamefont {Zaccone}\ and\ \citenamefont {Trachenko}(2020)}]{PNAS2020}%
  \BibitemOpen
  \bibfield  {author} {\bibinfo {author} {\bibfnamefont {A.}~\bibnamefont {Zaccone}}\ and\ \bibinfo {author} {\bibfnamefont {K.}~\bibnamefont {Trachenko}},\ }\href {\doibase 10.1073/pnas.2010787117} {\bibfield  {journal} {\bibinfo  {journal} {Proceedings of the National Academy of Sciences}\ }\textbf {\bibinfo {volume} {117}},\ \bibinfo {pages} {19653} (\bibinfo {year} {2020})},\ \Eprint {http://arxiv.org/abs/https://www.pnas.org/doi/pdf/10.1073/pnas.2010787117} {https://www.pnas.org/doi/pdf/10.1073/pnas.2010787117} \BibitemShut {NoStop}%
\bibitem [{\citenamefont {Li}\ and\ \citenamefont {Riedo}(2008)}]{Riedo}%
  \BibitemOpen
  \bibfield  {author} {\bibinfo {author} {\bibfnamefont {T.-D.}\ \bibnamefont {Li}}\ and\ \bibinfo {author} {\bibfnamefont {E.}~\bibnamefont {Riedo}},\ }\href {\doibase 10.1103/PhysRevLett.100.106102} {\bibfield  {journal} {\bibinfo  {journal} {Phys. Rev. Lett.}\ }\textbf {\bibinfo {volume} {100}},\ \bibinfo {pages} {106102} (\bibinfo {year} {2008})}\BibitemShut {NoStop}%
\bibitem [{\citenamefont {Noirez}\ and\ \citenamefont {Baroni}(2012)}]{Noirez}%
  \BibitemOpen
  \bibfield  {author} {\bibinfo {author} {\bibfnamefont {L.}~\bibnamefont {Noirez}}\ and\ \bibinfo {author} {\bibfnamefont {P.}~\bibnamefont {Baroni}},\ }\href {\doibase 10.1088/0953-8984/24/37/372101} {\bibfield  {journal} {\bibinfo  {journal} {Journal of Physics: Condensed Matter}\ }\textbf {\bibinfo {volume} {24}},\ \bibinfo {pages} {372101} (\bibinfo {year} {2012})}\BibitemShut {NoStop}%
\bibitem [{\citenamefont {Yang}\ \emph {et~al.}(2023)\citenamefont {Yang}, \citenamefont {Ji}, \citenamefont {Choi}, \citenamefont {Park}, \citenamefont {An}, \citenamefont {Lee}, \citenamefont {Jeong}, \citenamefont {Park}, \citenamefont {Kim}, \citenamefont {Park}, \citenamefont {Jeong}, \citenamefont {Kim},\ and\ \citenamefont {Park}}]{yang2023suppressed}%
  \BibitemOpen
  \bibfield  {author} {\bibinfo {author} {\bibfnamefont {H.}~\bibnamefont {Yang}}, \bibinfo {author} {\bibfnamefont {G.}~\bibnamefont {Ji}}, \bibinfo {author} {\bibfnamefont {M.}~\bibnamefont {Choi}}, \bibinfo {author} {\bibfnamefont {S.}~\bibnamefont {Park}}, \bibinfo {author} {\bibfnamefont {H.}~\bibnamefont {An}}, \bibinfo {author} {\bibfnamefont {H.-T.}\ \bibnamefont {Lee}}, \bibinfo {author} {\bibfnamefont {J.}~\bibnamefont {Jeong}}, \bibinfo {author} {\bibfnamefont {Y.~D.}\ \bibnamefont {Park}}, \bibinfo {author} {\bibfnamefont {K.}~\bibnamefont {Kim}}, \bibinfo {author} {\bibfnamefont {N.}~\bibnamefont {Park}}, \bibinfo {author} {\bibfnamefont {J.}~\bibnamefont {Jeong}}, \bibinfo {author} {\bibfnamefont {D.-S.}\ \bibnamefont {Kim}}, \ and\ \bibinfo {author} {\bibfnamefont {H.-R.}\ \bibnamefont {Park}},\ }\href@noop {} {\enquote {\bibinfo {title} {Suppressed terahertz dynamics of water confined in nanometer gaps},}\ } (\bibinfo {year} {2023}),\ \Eprint {http://arxiv.org/abs/2310.19236} {arXiv:2310.19236
  [physics.optics]} \BibitemShut {NoStop}%
\bibitem [{\citenamefont {Yu}\ \emph {et~al.}(2022)\citenamefont {Yu}, \citenamefont {Yang}, \citenamefont {Baggioli}, \citenamefont {Phillips}, \citenamefont {Zaccone}, \citenamefont {Zhang}, \citenamefont {Kajimoto}, \citenamefont {Nakamura}, \citenamefont {Yu},\ and\ \citenamefont {Hong}}]{Yu_2022}%
  \BibitemOpen
  \bibfield  {author} {\bibinfo {author} {\bibfnamefont {Y.}~\bibnamefont {Yu}}, \bibinfo {author} {\bibfnamefont {C.}~\bibnamefont {Yang}}, \bibinfo {author} {\bibfnamefont {M.}~\bibnamefont {Baggioli}}, \bibinfo {author} {\bibfnamefont {A.~E.}\ \bibnamefont {Phillips}}, \bibinfo {author} {\bibfnamefont {A.}~\bibnamefont {Zaccone}}, \bibinfo {author} {\bibfnamefont {L.}~\bibnamefont {Zhang}}, \bibinfo {author} {\bibfnamefont {R.}~\bibnamefont {Kajimoto}}, \bibinfo {author} {\bibfnamefont {M.}~\bibnamefont {Nakamura}}, \bibinfo {author} {\bibfnamefont {D.}~\bibnamefont {Yu}}, \ and\ \bibinfo {author} {\bibfnamefont {L.}~\bibnamefont {Hong}},\ }\href {\doibase 10.1038/s41467-022-31349-6} {\bibfield  {journal} {\bibinfo  {journal} {Nature Communications}\ }\textbf {\bibinfo {volume} {13}},\ \bibinfo {pages} {3649} (\bibinfo {year} {2022})}\BibitemShut {NoStop}%
\bibitem [{\citenamefont {Parker}\ \emph {et~al.}(2002)\citenamefont {Parker}, \citenamefont {Maria},\ and\ \citenamefont {Kingon}}]{Parker}%
  \BibitemOpen
  \bibfield  {author} {\bibinfo {author} {\bibfnamefont {C.~B.}\ \bibnamefont {Parker}}, \bibinfo {author} {\bibfnamefont {J.-P.}\ \bibnamefont {Maria}}, \ and\ \bibinfo {author} {\bibfnamefont {A.~I.}\ \bibnamefont {Kingon}},\ }\href {\doibase 10.1063/1.1490148} {\bibfield  {journal} {\bibinfo  {journal} {Applied Physics Letters}\ }\textbf {\bibinfo {volume} {81}},\ \bibinfo {pages} {340} (\bibinfo {year} {2002})}\BibitemShut {NoStop}%
\bibitem [{\citenamefont {Kittel}(1996)}]{Kittel}%
  \BibitemOpen
  \bibfield  {author} {\bibinfo {author} {\bibfnamefont {C.}~\bibnamefont {Kittel}},\ }\href@noop {} {\emph {\bibinfo {title} {Introduction to Solid State Physics, 7th edition}}}\ (\bibinfo  {publisher} {John Wiley and Sons, Inc.},\ \bibinfo {year} {1996})\BibitemShut {NoStop}%
\bibitem [{\citenamefont {Sidorova}\ \emph {et~al.}(2023)\citenamefont {Sidorova}, \citenamefont {Semenov}, \citenamefont {Charaev}, \citenamefont {Gonzalez}, \citenamefont {Schilling}, \citenamefont {Gyger},\ and\ \citenamefont {Steinhauer}}]{Sidorova_2023}%
  \BibitemOpen
  \bibfield  {author} {\bibinfo {author} {\bibfnamefont {M.}~\bibnamefont {Sidorova}}, \bibinfo {author} {\bibfnamefont {A.~D.}\ \bibnamefont {Semenov}}, \bibinfo {author} {\bibfnamefont {I.}~\bibnamefont {Charaev}}, \bibinfo {author} {\bibfnamefont {M.}~\bibnamefont {Gonzalez}}, \bibinfo {author} {\bibfnamefont {A.}~\bibnamefont {Schilling}}, \bibinfo {author} {\bibfnamefont {S.}~\bibnamefont {Gyger}}, \ and\ \bibinfo {author} {\bibfnamefont {S.}~\bibnamefont {Steinhauer}},\ }\href@noop {} {\enquote {\bibinfo {title} {Phonon heat capacity and disorder: new opportunities for performance enhancement of superconducting devices},}\ } (\bibinfo {year} {2023}),\ \Eprint {http://arxiv.org/abs/2308.12090} {arXiv:2308.12090 [cond-mat.supr-con]} \BibitemShut {NoStop}%
\bibitem [{\citenamefont {Guo}\ \emph {et~al.}(2022)\citenamefont {Guo}, \citenamefont {Zhang}, \citenamefont {Lu}, \citenamefont {Bai}, \citenamefont {Wen},\ and\ \citenamefont {Wang}}]{Wei-Hua}%
  \BibitemOpen
  \bibfield  {author} {\bibinfo {author} {\bibfnamefont {Q.}~\bibnamefont {Guo}}, \bibinfo {author} {\bibfnamefont {H.~P.}\ \bibnamefont {Zhang}}, \bibinfo {author} {\bibfnamefont {Z.}~\bibnamefont {Lu}}, \bibinfo {author} {\bibfnamefont {H.~Y.}\ \bibnamefont {Bai}}, \bibinfo {author} {\bibfnamefont {P.}~\bibnamefont {Wen}}, \ and\ \bibinfo {author} {\bibfnamefont {W.~H.}\ \bibnamefont {Wang}},\ }\href {\doibase 10.1063/5.0103336} {\bibfield  {journal} {\bibinfo  {journal} {Applied Physics Letters}\ }\textbf {\bibinfo {volume} {121}},\ \bibinfo {pages} {142204} (\bibinfo {year} {2022})},\ \Eprint {http://arxiv.org/abs/https://pubs.aip.org/aip/apl/article-pdf/doi/10.1063/5.0103336/16484582/142204\_1\_online.pdf} {https://pubs.aip.org/aip/apl/article-pdf/doi/10.1063/5.0103336/16484582/142204\_1\_online.pdf} \BibitemShut {NoStop}%
\bibitem [{\citenamefont {Baggioli}\ and\ \citenamefont {Zaccone}(2019)}]{Baggioli}%
  \BibitemOpen
  \bibfield  {author} {\bibinfo {author} {\bibfnamefont {M.}~\bibnamefont {Baggioli}}\ and\ \bibinfo {author} {\bibfnamefont {A.}~\bibnamefont {Zaccone}},\ }\href {\doibase 10.1103/PhysRevLett.122.145501} {\bibfield  {journal} {\bibinfo  {journal} {Phys. Rev. Lett.}\ }\textbf {\bibinfo {volume} {122}},\ \bibinfo {pages} {145501} (\bibinfo {year} {2019})}\BibitemShut {NoStop}%
\bibitem [{\citenamefont {Sidorova}\ \emph {et~al.}(2020)\citenamefont {Sidorova}, \citenamefont {Semenov}, \citenamefont {H\"ubers}, \citenamefont {Ilin}, \citenamefont {Siegel}, \citenamefont {Charaev}, \citenamefont {Moshkova}, \citenamefont {Kaurova}, \citenamefont {Goltsman}, \citenamefont {Zhang},\ and\ \citenamefont {Schilling}}]{Sidorova_2020}%
  \BibitemOpen
  \bibfield  {author} {\bibinfo {author} {\bibfnamefont {M.}~\bibnamefont {Sidorova}}, \bibinfo {author} {\bibfnamefont {A.}~\bibnamefont {Semenov}}, \bibinfo {author} {\bibfnamefont {H.-W.}\ \bibnamefont {H\"ubers}}, \bibinfo {author} {\bibfnamefont {K.}~\bibnamefont {Ilin}}, \bibinfo {author} {\bibfnamefont {M.}~\bibnamefont {Siegel}}, \bibinfo {author} {\bibfnamefont {I.}~\bibnamefont {Charaev}}, \bibinfo {author} {\bibfnamefont {M.}~\bibnamefont {Moshkova}}, \bibinfo {author} {\bibfnamefont {N.}~\bibnamefont {Kaurova}}, \bibinfo {author} {\bibfnamefont {G.~N.}\ \bibnamefont {Goltsman}}, \bibinfo {author} {\bibfnamefont {X.}~\bibnamefont {Zhang}}, \ and\ \bibinfo {author} {\bibfnamefont {A.}~\bibnamefont {Schilling}},\ }\href {\doibase 10.1103/PhysRevB.102.054501} {\bibfield  {journal} {\bibinfo  {journal} {Phys. Rev. B}\ }\textbf {\bibinfo {volume} {102}},\ \bibinfo {pages} {054501} (\bibinfo {year} {2020})}\BibitemShut {NoStop}%
\bibitem [{\citenamefont {Song}\ \emph {et~al.}(2004)\citenamefont {Song}, \citenamefont {Cui}, \citenamefont {Xia},\ and\ \citenamefont {Chen}}]{Song_linear}%
  \BibitemOpen
  \bibfield  {author} {\bibinfo {author} {\bibfnamefont {Q.}~\bibnamefont {Song}}, \bibinfo {author} {\bibfnamefont {Z.}~\bibnamefont {Cui}}, \bibinfo {author} {\bibfnamefont {S.}~\bibnamefont {Xia}}, \ and\ \bibinfo {author} {\bibfnamefont {S.}~\bibnamefont {Chen}},\ }\href {\doibase https://doi.org/10.1016/j.mejo.2004.06.013} {\bibfield  {journal} {\bibinfo  {journal} {Microelectronics Journal}\ }\textbf {\bibinfo {volume} {35}},\ \bibinfo {pages} {817} (\bibinfo {year} {2004})},\ \bibinfo {note} {therminic'03}\BibitemShut {NoStop}%
\bibitem [{\citenamefont {Zaccone}(2023)}]{zaccone2023}%
  \BibitemOpen
  \bibfield  {author} {\bibinfo {author} {\bibfnamefont {A.}~\bibnamefont {Zaccone}},\ }\href@noop {} {\emph {\bibinfo {title} {Theory of Disordered Solids}}}\ (\bibinfo  {publisher} {Springer},\ \bibinfo {address} {Cham},\ \bibinfo {year} {2023})\BibitemShut {NoStop}%
\end{thebibliography}%

\end{document}